\documentstyle[12pt]{article}
\newcommand{\be}{\begin{equation}}
\newcommand{\ee}{\end{equation}}
\newcommand{\bea}{\begin{eqnarray}}
\newcommand{\eea}{\end{eqnarray}}
\newcommand{\ba}{\begin{array}{l}}
\newcommand{\ea}{\end{array}}
\newcommand{\bb}{\begin{thebibliography}}
\newcommand{\eb}{\end{thebibliography}}
\newcommand{\ci}[1]{\cite{#1}}

\newcommand{\Tr}{\mbox{Tr\,}}
\newcommand{\Ds}{\displaystyle}

\title{TWO-LOOP EFFECTIVE ACTION\\
        FOR THEORIES WITH FERMIONS}

\author{B.A.FAIZULLAEV\thanks{Permanent address: Theoretical
                              Physics Department, Tashkent State
                                University, Tashkent 700095,
              Uzbekistan}\\ Physics Department, Middle East
                          Technical University\\ Ankara 06531,
                           Turkey\\ and\\
 M.M.MUSAKHANOV \\
  Theoretical Physics Department, Tashkent State University \\
     Tashkent 700095, Uzbekistan}

\begin{document}
\begin{titlepage}
\maketitle
      \begin{abstract}
On the basis of a new approach
proposed in our previous work we develope
a formalism for calculating of the effective action
for some models containing fermion fields.
This method allows us to calculate the fermionic part of
the effective action (up to two-loop level) properly.
The two-loop contribution to the effective
potential for the Nambu-Jona-Lasinio
model is calculated and is shown to vanish.
\end{abstract}
\vskip 50mm
Number of pages: 25.
Number of figures: 4.
\end{titlepage}
Running head: Two--loop effective action for theories with
fermions\\
Name and the mailing address of the author to whom proofs should
be sent:\\
Prof.Dr.Yousuf M.Musakhanov\\
Tashkent State University\\
Department of Theoretical Physics\\
700095     Tashkent, Uzbekistan.\\
tel: 7 3712 46 52 64\\
fax: 7 3712 39 45 16 \\
E-Mail: yousuf@iceeat.silk.glas.apc.org
\hskip 0.5cm
yousuf@univer.tashkent.su\\
\newpage
\section{Introduction}
The functional formulations of quantum field theories are the
most attractive ones for investigation of general properties of
these theories. The complete quantum-theoretical information can
be extracted from the effective action functional. The effective
action generates all 1PI Green's functions. But there are no
special methods for the calculation of the effective action and
usually the effective action is calculated either by direct
summation of the infinite series of the Feynman diagrams \ci{1}
or by the functional integration method \ci{2,3}. In our previous
works \ci{4} we have developed a new method for the calculation
of the effective action for quantum field models. This method is
based on an old formula of DeWitt \ci{5} that connects the vacuum
expectation of a quantum functional with the classical ones ( and
its derivatives with respect to the classical fields ).

 The application of DeWitt formula leads to variational
differential equations for effective action; by iterations over
Planck constant $\hbar$ we can derive from them a loop expansion
( for scalar theory this was done in \ci{4}). Such computational
method of course is simpler and more convenient compared to other
 methods \ci{1,2,3} -- we don't deal with the coefficients of
each class of diagrams or with the picking out of nonacceptable
diagrams as in \ci{3}.

By this method we have calculated the two - loop effective action
for scalar $\lambda\varphi^4$ theory and for spinor
electrodynamics in \ci{4}. But in the last case our answer was
not complete - we had not calculated fermionic part of the
effective action. This work will be devoted to calculation of the
complete effective action including the fermionic part of them
for some fermionic models , namely, for QED and for bosonized
Nambu-Jona-Lasinio ( NJL ) model. Fermionic part of the effective
action is the part that generates amplitudes of processes with
fermions at initial and/or final states.

The NJL model has a very attractive property: it manifests the
dynamical acquiring of mass by fermion caused by spontaneous
symmetry breaking \ci{6,7}. For investigation of the spontaneous
symmetry breaking it is necessary to know the effective potential
of the theory. The effective potential may be derived from the
effective action by setting all the classical fields equals to
constants. There are many works devoted to the calculation of the
effective potential for different versions of the NJL model (see
\ci{8} and references therein), but usually almost all authors
are dealing with one-loop approximation or some asymptotics of
the potential. In this work we calculate the effective potential
for the NJL model at the two-loop level for two cases: two- and
four-dimensional space-time. We show that the two-loop
contribution to the effective potential vanishes for the both
cases. Thus, the dynamical spontaneous symmetry breaking property
of the NJL model is determined only by the one-loop correction.

The article is organized as follows: in Sec.2 we derive the
DeWitt formula for scalar and for spinor theories; in Sec.3 we
describe the relationships between the classical and the
effective actions; in Sec.4 we calculate the two--loop effective
action for scalar $\lambda\varphi^4$ theory and in Sec.5 we
describe a formal method of integrating of the equations. In
Sec.6 we introduce a new entity-the functional which is the
effective action involving only one variable and is the
generating functional of the connected Green's functions with
respect to another variables. This functional is analogous to the
well known Rauss function in classical dynamics. Introduction of
this functional permits us to calculate amplitudes of any
processes with fermions at initial and/or final states. In the
remaining part of the Sec.6 we calculate the two--loop effective
action for QED and in Sec.7 the two-loop effective action for the
bosonized NJL model. In Sec.8 we will show that the two--loop
contribution to the effective potential of the NJL model
vanishes. \section{DeWitt's formula}

Let's begin with scalar case and define the generating functional
of Green's functions as: \be Z[J]=\exp\left(iW[J]\right)=\int
D\Phi\exp\left(i\int\left(L+J\Phi\right)dx\right), \ee where $
W[J] $ is generating functional of the connected Green's function
and $ L $ is is the Lagrangian for scalar field $\Phi(x)$.
Consider the following expression: \be \ba
\Ds{\exp\left(iW[J+\zeta]\right)=\int D\Phi\exp\lbrace
i\int\left(L+(J+\zeta)\Phi \right)dx\rbrace=}\\ \\
\Ds{=\sum_{n=0}^{\infty}\frac{i^n}{n!}\zeta_{i_{1}}\zeta_{i_{2}}\cdots
\zeta_{i_{n}} {\int}D\Phi
\Phi_{i_{1}}\Phi_{i_{2}}\ldots\Phi_{i{n}}\exp\lbrace i
\int\left(\it L+J\Phi\right){dx}\rbrace,} \ea \ee where we use
the condensed notations $\Phi_{i}=\Phi(x_{i}),
\Phi_{i}\zeta_{i}=\int\Phi(x)\zeta(x)dx$. If $\hat\varphi$ is a
quantum field then we may define vacuum expectation value of any
quantum functional $Q[\hat\varphi]$ as follows $$\langle
Q[\hat\varphi]\rangle =\exp\left({\em -i}W[J]\right)
\int{D\Phi}Q[\Phi]\exp\left({\em i}\int\left({\em
 L}+J\Phi\right)dx\right).$$ In this formula $Q[\Phi]$ in the
integrand is the classical functional, that is the functional
depending on c-functions $\Phi(x)$. Let's expand $Q[\Phi]$ in the
integrand in the Taylor series: \be \ba \Ds{\langle
Q[\hat\varphi]\rangle=\exp\left(-iW[J]\right)\sum^{\infty}_{n=0}
{1\over n!}\frac{\delta^n W}{\delta \Phi^i_{1}\delta
\Phi^i_{2}\cdots \delta \Phi^i_{n}}_{\Phi=0}\cdot}\\ \\
\Ds{{\int}D\Phi
\Phi_{i_{1}}\Phi_{i_{2}}\ldots\Phi_{i{n}}\exp\lbrace i
\int\left(\it L+J\Phi\right){dx}\rbrace,} \ea \ee Comparing
Eq.(2) and Eq.(3) we get: \be \langle
Q[\hat\varphi]\rangle=:\exp\left(-iW[J]\right)
\exp\left(iW[J+\frac{\delta}
{i\delta\Phi}]\right):Q[\Phi]_{\Phi=0},\ee where the colons mean
that derivatives acts only on $ Q[\Phi]$. Substituting the
following expansion $$ W[J+\zeta]=W[J]+\zeta_i\frac{\delta
W}{\delta J_i}+\sum^{\infty}_{n=2}
\frac{1}{n!}\zeta_{i_{1}}\zeta_{i_{2}}\cdots\zeta_{i_{n}}
G^{i_1i_2\cdots i_n},$$ where $$ G^{i_1i_2\cdots
i_n}=\frac{\delta^nW}{\delta J_{i_1}\delta J_{i_2}\cdots\delta
J_{i_n}} $$ is connected Green's functions, into Eq.(4) we get
the following formula of DeWitt (\ci{5}, ch.22):
   \be
\langle{Q[\hat\varphi]}\rangle=:\exp\left(\frac{i}{\hbar}
\sum_{n=2}^{\infty}\frac{(-i\hbar)^n}{n!}
G^{i_{1}i_{2}\ldots i_{n}}\frac{\delta^{n}}{\delta\varphi^{i_1}
\delta\varphi^{i_2}\ldots\delta\varphi^{i_n}}\right):Q[\varphi]
\ee where the $\varphi$ in the rhs of this equation means the
so-called classical fields: \be \varphi_{\em
i}={\delta}W[J]/{\delta}J_i.\ee
  Since we will deal with the loop expansion we have restored
the Planck constant $\hbar$ in DeWitt's formula.

For the models containing fermion and vector fields
the generalization of DeWitt's formula leads to the following
expression:
\be \ba \langle {Q[\hat A_{\mu},\hat\psi,\hat{\bar\psi}]}\rangle
 =:\exp  \left( \hat G\right):
 Q[A_{\mu},\psi,\bar\psi]
\ea
\ee
where the operator $\hat G$  is defined as follows:
$$
\ba
\Ds{\hat G= \frac {i}{\hbar}\sum_{n=2}^{\infty }
 \frac{(-i\hbar )^n}{n!}
 \sum C^{n}_{ijk}(-1)^{j}\cdot }\\   \\
\Ds{\cdot G^{\mu_1\ldots\mu_i\alpha_1\ldots\alpha_j\beta_1
\ldots\beta_k}
\frac {\delta^n}{\delta A_{\mu_1}\ldots \delta A_{\mu_i}
 \delta\psi_{\beta_k}\ldots\delta\psi_{\beta_1}
 \delta\bar\psi_{\alpha_j}\ldots\delta\bar\psi_{\alpha_1}}.}
 \ea $$
 The classical fields are defined as follows:
 $$A_{\mu}=\delta W/\delta J_\mu,\qquad \psi =\delta W
/\delta\bar\eta ,\qquad \bar\psi =-\delta W/\delta\eta.$$ The
coefficients $ C^{n}_{ijk} $ are defined as $$
C^{n}_{ijk}=n!/(i!j!k!),\qquad i+j+k=n, $$ and
$$G^{\mu_1\ldots\mu_i\alpha_1\ldots\alpha_j\beta_1\ldots\beta_k}=
 \delta^{n} {W}/\left({\delta}J_{\mu_1}\ldots{\delta}J_{\mu_i}
 \delta\eta_{\beta_{k}}\ldots\delta\eta_{\beta_{1}}
 \delta\bar\eta_{\alpha_j}\ldots\delta\bar\eta_{\alpha_1}\right)$$
- are the connected Green's function with $i$ photon and $j+k$
fermion legs. Therefore, the vacuum expectation value of a
quantum functional is defined by the (classical) functional and
derivatives of them where quantum fields are replaced by
classical fields.
\section{The Relations Between Classical and Effective Actions}
Usually the effective action $\Gamma[\varphi]$ is defined as
\be\Gamma[\varphi]=W[J]-J_i\varphi^i, \ee
where $\varphi_i=\delta W[J]/\delta J_i$; but for our purposes a
 more convenient definition is that of DeWitt \ci{5}:
\begin{equation} \langle\frac{\delta
S}{\delta\hat\varphi_i}\rangle=
\frac{\delta\Gamma}{\delta\varphi_{\em i}}. \end{equation} This
definition is a consequence of the following well known
relations: \be \frac{\delta\Gamma}{\delta\varphi^i}=-J_i,\qquad
\langle\frac{\delta S}{\delta \hat\varphi^i}\rangle=-J_i.\ee The
last of these is the "quantum equations of motion" and the first
is consequence of Eq.(8). Comparing of Eq.(5) and Eq.(9) leads to
the main formula: \be
\frac{\delta\Gamma}{\delta\varphi^i}=:\exp\lbrace\frac{i}{\hbar}
\sum^{\infty}_{n=2}
\frac{(-i\hbar)^n}{n!}G^{i_1i_2\cdots
i_n}\frac{\delta^n}{\delta\varphi^i_{1}
\delta\varphi^i_{2}\cdots\delta\varphi^i_{n}}\rbrace:\frac{\delta
S}{\delta\varphi^i}. \ee This formula in principle, permits us by
using the known classical action $ S $ to restore the quantum
effective action $\Gamma $. We need only an additional relation
between $\Gamma $ and $ G^{ij} $. Clearly this relation and
Eq.(11) will have different forms for different theories.

For the spinor electrodynamics
\be
 L=-\frac{1}{4}F_{\mu\nu}F^{\mu\nu}+\bar\Psi \left[i\left(\hat
\partial - ie\hat A\right ) +m\right]\Psi -\frac{1}{2\alpha}\left
(\partial _{\mu}A^{\mu}\right )^2 \ee we may write relations
connecting classical and effective actions as follows:
\begin{equation} \frac{\delta\Gamma}{{\delta}A_{\mu}} =
\langle\frac{{\delta}S}{\delta\hat A_{\mu}}\rangle,\qquad
\frac{\delta\Gamma}{\delta{\bar\psi}^\alpha} =
\langle\frac{{\delta}S}{\delta\hat{\bar\psi}^\alpha}\rangle,
\qquad \frac{\delta\Gamma}{\delta{\psi}^\alpha} =
\langle\frac{{\delta}S}{\delta\hat{\psi}^\alpha}\rangle.
\end{equation}
 These formulas are most convenient for application of Eqs.(7)
 and (9).
\section{Effective Action for Scalar $\lambda\varphi^4/4!$ Theory}
Let's take the classical action in the following form:
\be S=
-\frac{1}{2}\varphi(\partial^2+m^2)\varphi-\lambda\varphi^4/4!.
\ee Since $$ \frac{\delta
S}{\delta\varphi^i}=-(\partial^2+m^2)\varphi_i-
\lambda\varphi^3_i/6, $$
then in Eq.(11) leaves finite number of terms: \be
\frac{\delta\Gamma}{\delta\varphi^i}=
iK^{-1}_{ij}\varphi^j-\frac{\lambda}{6}\varphi^3_i
+\frac{1}{2}i\lambda\hbar
G^{ii}\varphi^i+\frac{\lambda}{6}\hbar^2G^{ij}G^{ik}
G^{il}\frac{\delta^3\Gamma}{\delta\varphi^j
\delta\varphi^k\delta\varphi^l}.
\ee
This is (variational) differential equations for the
effective action $\Gamma[\varphi] $ \ci{4}. Here $$
K_{ij}=-i(\partial^2+m^2)^{-1}\delta_{ij} $$ is the free Feynman
Green's function and in third and fourth terms there are no
summations over $ i $. Differentiating the first of Eq.(10) with
respect to $J_k $ we get
\be
G^{ij}\frac{\delta^2\Gamma}{\delta\varphi^j\delta\varphi^k}=-
\delta^i_k. \ee
Equations (15) and (16) forms the system of equations for the
calculation of the effective action for the scalar theory. We
would like to derive the loop expansion; thus we will iterate
equations (15) and (16) over $\hbar $.

Let's expand $\Gamma $ and $ G $ over $\hbar $:
\be  \ba
\Gamma=\Gamma_0+\hbar\Gamma_1+\hbar^2\Gamma_2+\cdots,\\ \\
G=G_0+\hbar G_1+\hbar^2 G_2+\cdots,
\ea \ee
where $\Gamma_0, \Gamma_1, \Gamma_2,\ldots $--corresponds to the
tree, the one--loop, the two--loop , etc., approximations,
respectively. This is also true for $ G_0, G_1, G_2,\ldots $,
etc.,respectively. The substitution of Eq.(17) to Eq.(16) permits
us rewrite Eq.(16) in a more convenient form for the loop
expansion: \be
  G^{ij}_0 \frac{\delta^2 \Gamma_0}{\delta \varphi^j\delta
  \varphi^k}=
 -\delta^i_k ,\qquad G^{ij}_1 \frac{\delta^2\Gamma_0}{\delta
  \varphi^j\delta \varphi^k}+ G^{ij}_0 \frac{\delta^2
  \Gamma_1}{\delta \varphi^j\delta \varphi^k}=0, \ldots\ee From
   Eq.(15) we have the following equation for $\Gamma_0$: \be
\frac{\delta\Gamma_0}{\delta\varphi^i}=iK^{-1}_{ij}\varphi^j-
\frac{\lambda}{6}
\varphi^3_i. \ee This equation gives us: \be
\Gamma_0=\frac{1}{2}i\varphi^iK^{-1}_{ij}\varphi^j-\frac
{\lambda}{24}(\varphi^i)^4,
\ee or, in the usual notations
$$\Gamma_0=\frac{i}{2}\int\varphi(x)K^{-1}(x-y)\varphi(y)dxdy
\frac{\lambda}{4!}\int\varphi^4(x)dx=S.$$
It is suitable to introduce the following notation: \be
\Phi_{ij}=iK_{ij}^{-1}-\frac{\lambda}{2}(\varphi^i)^2\delta_{ij}=
iK^{-1}_{ik}\left(1+i\frac{\lambda}{2}K\varphi^2\right)_{kj}. \ee
Then from the first of Eqs.(18) we get: \be
G_0^{ij}=-(\Phi^{-1})^{ij}. \ee Now we may to calculate the
$\Gamma_1 $. First note that: \be
\frac{\delta\Gamma_1}{\delta\varphi^i}=i\frac{\lambda}{2}G_0^{ii}
\varphi^i=-i\frac{\lambda}{2}(\Phi^{-1})^{ii}\varphi^i,
\ee
which can be rewritten as follows:
\be
\delta\Gamma_1=\frac{i}{2}(\Phi^{-1})^{ij}\delta\Phi_{ji}.
\ee
The integration of Eq.(24) gives us the well known one--loop
result (after the second equality sign we are dropping the
trivial infinite term): \be
\Gamma_1=\frac{i}{2}\Tr\log\Phi=\frac{i}{2}\Tr\log\left(1+
i\frac{\lambda}{2}K\varphi^2\right).
\ee We get for $\Gamma_2 $ : $$
\Ds{\frac{\delta\Gamma_2}{\delta\varphi^i}=
i\frac{\lambda}{2}G_1^{ii}\varphi^i+
\frac{\lambda}{6}G_0^{ij}G_0^{ij}G_0^{ik}G_0^{il}
\frac{\delta^3\Gamma_0}
{\delta\varphi^j\delta\varphi^k\delta\varphi^l}} $$ (on the
r.h.s. of the equation there are no summations over $ i $).
Determining $ G_1 $ from the second of the Eq.(18) we find: $$
\ba
\Ds{\frac{\delta\Gamma_2}{\delta\varphi^i}=\frac{\lambda^2}{4}
(\Phi^{-1})^{is}(\Phi^{-1})^{ss}(\Phi^{-1})^{si}\varphi^i+}\\ \\
\Ds{+\frac{\lambda^3}{4}(\Phi^{-1})^{il}(\Phi^{-1})^{kl}\varphi^l
(\Phi^{-1})^{lk}\varphi^k(\Phi^{-1})^{ki}\varphi^i+
\frac{\lambda^2}{6}((\Phi^{-1})^{il})^3\varphi^l} \ea $$ (there
are no summations over $ i $). After some manipulations we get
\ci{4}: \be
\Gamma_2=\frac{\lambda}{8}((\Phi^{-1})^{ii})^2+\frac{\lambda^2}{12}
 \varphi^i((\Phi^{-1})^{ij})^3\varphi^j. \ee To this expression
corresponds the set of diagrams depicted in Fig.1. \vskip 20mm
\centerline{Figure 1.}
\section{Effective action for QED}
Application of our main Eq.(7) to Eqs.(13) give rise to
the following system of equations :
\be
\frac{\delta\Gamma}{\delta A_{\mu}} = e\bar\psi\gamma^\mu\psi+
D^{-1\mu\nu}A_\nu+ie\hbar \Tr\left(\gamma^\mu\frac{\delta^2W}
{\delta\eta\delta\bar\eta}\right),
\ee
\be
\frac{\delta\Gamma}{\delta\bar\psi^ \alpha} = \left ((i\hat \partial+
e\hat A-m)\psi\right)^\alpha- \it ie\hbar\left(\gamma_{\mu}\right)^
{\alpha\beta}\frac{\delta^2W}{\delta J_{\mu}\delta\bar\eta^{\beta}},
\ee
\be
\frac{\delta\Gamma}{\delta\psi^ \alpha} = \left (\bar\psi(i
\overleftarrow
{\hat\partial}-e\hat A+m)\right)^\alpha- ie\hbar\left
(\gamma_{\mu}\right)^
{\beta\alpha}\frac{\delta^2W}{\delta J_{\mu}\delta\eta^{\beta}}.
\ee
where
$$ D^{\mu\nu}=\frac{1}{\partial^2}\left( g^{\mu\nu}-(1-\alpha)\frac
{\partial^{\mu}\partial^{\nu}}{\partial^2}\right) $$
is the free photon Green's function.

 The system of equations (27)-(28)-(29) is not complete - to
those we must add equations relating $\Gamma $ to the connected
Green's functions $ G $.
     For the spinor model the relations between $\Gamma$ and $G$
 are very complicated, and for this reason we developed a new
 formalism (which follows).

 Let's introduce the following notations for the classical fields
 and the sources: \be \alpha=(A_{\mu},-\psi,\bar\psi),\qquad
 \beta=(J_{\mu},\bar\eta,\eta), \ee and also \be
 \Gamma_{i_1i_2\cdots
  i_n}=\frac{\delta^n\Gamma}{\delta\alpha^{i_1}\delta\alpha^{i_2}
  \cdots \delta\alpha^{i_n}},
  \qquad G^{ij\cdots
n}=\epsilon_{i_{1}}\cdots\epsilon_{i_{n}}\frac{\delta^n
W}{\delta\beta_{i_1}\delta\beta_{i_2}\cdots \delta\beta_{i_n}},
\ee where $\epsilon_i$ is equal to $+1$ if $i=\mu$ and $-1$ if
otherwise. In these notations the index $ i $ may take the values
 $ \mu,\alpha $ and $\bar\alpha $, where $\alpha ( \bar\alpha)$
 correspond to the index of the field $\psi^{\alpha} (
 \bar\psi^{\alpha})$. Then our equations (27-28-29) can be
 written in the form: \be \Gamma_i=S_i+ie\hbar\gamma_{ijk}G^{kj}
\ee where \be \gamma_{ijk}
=\left\{\begin{array}{ll}\gamma_\mu^{\alpha\bar\beta}&{\rm if\ }
i=\mu,\\ \gamma_\mu^{\bar\beta\alpha}&{\rm if\ } i=\alpha,\\
\gamma_\mu^{\bar\alpha\beta}&{\rm if\ } i=\bar\alpha.
\end{array}\right. \ee In this notations the equation for
$\Gamma$ have the following form $$\Gamma_i=-\beta_i.$$
Differentiating it with respect to $\beta_j$ we may derive
relation between $\Gamma$ and $G's$: \be
G^{ij}\Gamma_{jk}=-\epsilon_i\delta^i_k .\ee

Now we may rewrite our equations as follows:
\be
 \Gamma_i=S_i-ie\hbar\gamma_{ijk}\epsilon_j\left
 (\hat\Gamma^{-1}\right)^{kj}
\ee
where $\hat\Gamma^{-1}$ is the inverse matrix of $\{\Gamma_{ij}\}$.
The solution of this system of equations (35) provide us the
effective action for QED. Next we will present a formal approach
for solving this system.

Expanding $\Gamma$ in terms of $\hbar$ as in Eq.(17)
we obtain for $\Gamma_0$:
$$\Gamma_{0i}=S_i.$$
That is, we have
\be \Gamma_0=S, \ee
as it must be. For $\Gamma_1$ we have
\be
\Gamma_{1i}=-ie\gamma_{ijk}\epsilon_j \left(\hat
S^{-1}\right)^{kj}. \ee In other words \be \delta\Gamma_{1}=
-ie\gamma_{ijk}\delta\alpha^{i}\epsilon_j \left(\frac {\delta^2
S}{\delta\alpha^{j}\delta\alpha^{k}}\right)^{-1}. \ee If we
return to usual notations we have: \be \ba
\delta\Gamma_{1}=-ie\delta
A^{\mu}\Tr\left[\gamma_{\mu}\left(\frac{\delta^2
S}{\delta\psi\delta\bar\psi}\right)^{-1}\right]+ \\
+ie\left[\delta\bar\psi\gamma_{\mu}\left(\frac{\delta^2
S}{\delta\bar\psi\delta A_{\mu}}\right)^{-1}
-\left(\frac{\delta^2 S}{\delta\psi\delta
A_{\mu}}\right)^{-1}\gamma_{\mu}\delta\psi\right]. \ea \ee The
first term of Eq.(39) can be easily integrated to yield: \be
\Gamma^{A}_{1}=-i\Tr\log\left(i\hat\partial+e\hat A-m\right). \ee
For the non-fermionic part of $\Gamma $ we can compute also the
two-loop contribution without serious difficulties \ci{4}. For
this we neglect all the terms with fermions in initial and/or
final states and get for non-fermionic part of the $\Gamma $: \be
\Ds{\frac{\delta\Gamma^{A}}{\delta A_\mu}}=D^{-1\mu\nu}A_{\nu}-
ie\hbar\Tr\left(\hat K\gamma^{\mu}\right)
-e^2\hbar^2\Tr\left(G^{\nu}\gamma_{\nu}\hat
K\gamma^{\mu}\right),\ee where $ G^{\nu} $ is three--point
connected Green's function with one photon and two electron legs
and $$\hat K^{-1}=i\hat\partial +e\hat A-m. $$ The last term of
Eq.(41) consist of two--, three--, and etc. loops contributions
to $\Gamma^{A}$. For example, the two--loops contribution to
$\delta\Gamma^{A}/\delta A_{\mu}$ has the following form: $$
e^3\hbar^2\Tr\left(\gamma^{\mu}\hat K\gamma_{\sigma}\hat
K\gamma_{\nu}\hat K\right)D^{\sigma\nu}, $$ from which we get: $$
\delta\Gamma^{A}_{2}=-e^2\Tr\left(\delta\hat K\gamma_{\sigma}\hat
K\gamma_{\nu}\right)D^{\sigma\nu}.$$ Hence, \be
\Gamma^{A}_2=-\frac{1}{2}e^2\Tr\left(\gamma_{\mu}\hat
K\gamma_{\nu}\hat K\right)D^{\mu\nu}.\ee The first term of this
expression ( if we expand $\hat K $ over $\hat A $ ) is the
vacuum polarization diagram, other terms contribute to the
amplitudes with only photons in the external legs.

But the calculation of the fermionic part of $\Gamma $ is a
rather difficult problem. Due to this, we will use another method
to solve this problem in the following section .
\section{Fermionic Part of the Effective Action}
 Let us define the generating functional
\be Z[J_\mu,\eta,\bar\eta]= \int{DA_\mu}D\psi D\bar\psi
\exp\left({i}\int\left({L}+JA_\mu+\bar\eta\psi+\bar\psi\eta
\right)dx\right).\ee The integrations over spinor variables leads
us to an effective generating functional: \be
Z[J_\mu,\eta,\bar\eta]=
\int{DA_\mu}\exp\left({i}S_{eff}\right),\ee where \be
S_{eff}=S_A- F+i\hbar T, \ee and: \be \ba S_A=\frac{1}{2}A_\mu
D^{-1}_{\mu\nu}A_\nu,\qquad
  F=\bar\eta \hat K\eta,\qquad T=\Tr\log\hat K,\\ \\ \Ds{\hat
K^{-1}=i\hat\partial+e\hat A -m,\qquad
D_{\mu\nu}=\frac{1}{\partial^2}\left(g_{\mu\nu}-\left(1-
\alpha\right)\frac{\partial_\mu \partial_\nu}{\partial^2}\right).}
\ea \ee Let's define the following "effective action" \be
\tilde\Gamma[A_\mu,\eta,\bar\eta]=W[J_\mu,\eta,\bar\eta]-J_\mu
A^\mu \ee which is the effective action in terms of the $A_\mu$
variable and generates connected Green's functions with respect
to $\eta$ and $\bar\eta$ variables. So, we perform Legendre
transformations only with respect to $J_{\mu}$ variables without
disturbing $ (\eta,\bar\eta) $ variables. The new functional
$\tilde\Gamma $ is analogous to the Rauss function in classical
mechanics. By the application of the DeWitt's formula to $
S_{eff}$ we get the following expression: \be
\frac{\delta\tilde\Gamma}{\delta A_\mu}= :\exp(\hat
G):\frac{\delta S_{eff}} {\delta A_\mu},\ee where \be\hat
G=\frac{-i\hbar}{2}G^{\mu\nu}\frac{\delta^2}{\delta A^\mu\delta
A^\nu}
-\frac{\hbar^2}{6} G^{\mu\nu\lambda}\frac{\delta^3}{\delta
A^\mu\delta A^\nu\delta A^\lambda} +\ldots,\ee and
$$G^{\mu\nu\ldots\lambda}= \frac{\delta^n W}{\delta A_\mu\delta
A_\nu\ldots\delta A_\lambda}.$$ After substituting the following
formula \be \frac{\delta S_{eff}}{\delta
A_\mu}=D^{-1\mu\nu}A_\nu+ F^\mu-i\hbar T^\mu ,\ee into Eq.(48),
and introducing \be F^{\mu\nu\ldots\lambda}=(-1)^s\frac{\delta^s
F}{\delta A_\mu\delta A_\nu\ldots\delta A_\lambda},\qquad
T^{\mu\nu\ldots\lambda}=(-1)^s\frac{\delta^s T}{\delta
A_\mu\delta A_\nu\ldots\delta A_\lambda} \ee we get \be \ba
\delta\tilde\Gamma/\delta A_\mu= D^{-1\mu\nu}A_\nu+ F^\mu-i\hbar
T^\mu
-\frac{i}{2}\hbar G^{\sigma\lambda}\left(F_{\sigma\lambda\mu}
-i\hbar T_{\sigma\lambda\mu}\right)- \\ \\
 -\hbar^2[\frac{1}{6}G^{\sigma\lambda\nu}\left(-F_{\sigma\lambda\nu\mu}
 +i\hbar T_{\sigma\lambda\nu\mu}\right)+\frac{1}{8}
 G^{\sigma\lambda}G^{\nu\rho} \left(
 F_{\sigma\lambda\nu\rho\mu}-i\hbar
 T_{\sigma\lambda\nu\rho\mu}\right)]+\ldots \ea \ee Expanding $G$
and $\tilde\Gamma$ in terms of $\hbar$ as in Eq.(17) we get: \be
\frac{\delta\tilde\Gamma_0}{\delta A_\mu}= D^{-1\mu\nu}A_\nu+
 F^\mu ,\ee which provides us \be \tilde\Gamma_0=\frac{1}{2}A_\mu
 D^{-1\mu\nu}A_\nu-\hat F = S. \ee Using \be G^{\mu\nu}
 \frac{\delta^2 \tilde \Gamma}{\delta A^\nu \delta A^\lambda}=
 -\delta^\mu_\lambda \ee or, in other words \be G^{\mu\nu}_0
 \frac{\delta^2 \tilde \Gamma_0}{\delta A^\nu \delta A^\lambda}=
 -\delta^\mu_\lambda ,\qquad G^{\mu\nu}_1 \frac{\delta^2 \tilde
  \Gamma_0}{\delta A^\nu \delta A^\lambda}+ G^{\mu\nu}_0
  \frac{\delta^2 \tilde \Gamma_1}{\delta A^\nu \delta
   A^\lambda}=0 \ldots\ee we get: $$ G^{\mu\nu}_0
 \left(D^{-1}_{\nu\lambda}-F_{\nu\lambda}\right)=-
 \delta^{\mu}_{\lambda},$$
   or, in the matrix notation \be \hat G_0 = - \hat D
  \left(I-\hat F \hat D\right)^{-1}. \ee The last matrix equation
  must be understand as follows: $$
  G^\mu_{0\nu}=-D^{\mu}_{\sigma}\left(
  \delta^\sigma_\nu+F^{\sigma\lambda}D_{\lambda\nu}+
  F^{\sigma\lambda} D_{\lambda\rho} F^{\rho\alpha}
  D_{\alpha\nu}+\ldots\right). $$ For $\tilde\Gamma_1$ we get the
 equation \be \ba \Ds{\frac{\delta \tilde \Gamma_1}{\delta
A_\mu}= -iT^\mu-\frac{i}{2}G^{\sigma\lambda}_0
  F^\mu_{\sigma\lambda}=i\frac{\delta}{\delta
  A_\mu}T+\frac{i}{2}G^{\sigma\lambda}_0 \frac{\delta}{\delta
A_\mu}F_{\sigma\lambda}=} \\ \\ \Ds{ =\frac{\delta}{\delta
 A_\mu}[iT+\frac{i}{2}\Tr\log(I-\hat F\hat D)] ,} \ea \ee which
  after integration gives us: \be \ba
  \tilde\Gamma_1=iT+\frac{i}{2}\Tr\log(I-\hat F\hat D)= \\ \\
  =i\Tr\log\hat K
  +\frac{i}{2}\Tr\log\left(1-e^2\left(\bar\eta\hat K\gamma_\mu
  \hat K\gamma_\nu \hat K\eta+\bar\eta\hat K\gamma_\nu \hat
  K\gamma_\mu\hat K\eta \right) D^{\mu\nu}\right). \ea \ee The
  second term of the Eq.(59) represents the one-loop fermionic
  part of the "effective action" $\tilde\Gamma$. By
  differentiation of $\tilde\Gamma$ we may compute any (
  one-loop) amplitudes with fermion legs.

  The equation for $\tilde\Gamma_2$ has the form \be \frac{\delta
  \tilde\Gamma_2}{\delta A_\mu}=-\frac{i}{2}G_1^{\sigma\lambda}
  F^\mu_{\sigma\lambda}-\frac{1}{2}
  G_0^{\sigma\lambda}T_{\sigma\lambda}^\mu
  +\frac{1}{6}G_0^{\sigma\lambda\nu}F_{\sigma\lambda\nu}^\mu \\
  -\frac{1}{8}G_0^{\sigma\lambda}G_0^{\nu\rho}
  F_{\sigma\lambda\nu\rho}^\mu
   \ee For solving this equation we must first calculate $G_1$
  from Eq.(56). As shown in Appendix A $ G_{1} $ is: $$
  G_1^{\mu\nu}=G_0^{\mu\sigma}\left(iT_{\sigma\lambda}-
  \frac{i}{2}\frac{\delta}
{\delta
A^\sigma}\left(G_0^{\alpha\beta}F^{\lambda}_{\beta\alpha}\right)
\right).$$
and then the solution of the Eq.(60) is \be \tilde
  \Gamma_2=\frac{1}{2}\Tr(\hat G_0 \hat T) + \frac{1}{8}
  G_0^{\alpha\beta} G_0^{\gamma\delta}
  F_{\alpha\beta\gamma\delta}\\
  -\frac{1}{12}G_0^{\alpha\sigma} G_0^{\beta\nu}
   G_0^{\gamma\lambda} F_{\alpha\beta\gamma}F_{\sigma\nu\lambda}
   \ee

   Each term of this expression may be represented in graphic
   form as in Figs.2,3 and 4. \vskip 20mm \centerline{Figure 2.}
\vskip 20mm \centerline{Figure 3.} \vskip 20mm \centerline{Figure
4.}

The expressions for $\tilde\Gamma_0, \tilde\Gamma_1,
  \tilde\Gamma_2$ of course, have a symbolic sense; we have
  determined the algebraic structure of the effective action and
  each term of these expressions is a complicated integral. For
  derivation of the explicit formulas for effective action, all
  the integrations and trace calculations must be performed. Such
  a program can be carried out with specified functions
  $A_{\mu},\psi $ etc. in integrands.
\section{ The Effective Action for the NJL model}
 We would like to calculate the effective action for one of the
 widely discussed models-the model of Nambu-Jona-Lasinio. We take
 the model in the following bosonized form: \be L=\bar\psi
(i\hat\partial+g(\sigma+i\gamma_5\pi))\psi-
\frac{1}{2}(\sigma^2+\pi^2) \ee
 where $\sigma,\pi$ are auxiliary scalar and pseudoscalar fields.

 The equations for effective action for this model are \be \ba
 \Ds{ \frac{\delta\Gamma}{\delta \phi^{i}}=
-\phi^{i}+g\bar\psi\gamma_{i}\psi+
 g\Tr\left(\gamma_{i}\frac{\delta^W}{\delta\eta\delta\bar\eta}
 \right), }\\ \\
\Ds{ \frac{\delta\Gamma}{\delta \psi}=
 \bar\psi(i\overleftarrow{\hat\partial}-g\hat\phi)-
ig\hbar\gamma_{i}\frac{\delta^2W}{\delta\eta\delta J_{i}},}\\ \\
 \Ds{ \frac{\delta\Gamma}{\delta \bar\psi}=
 (i\hat\partial+g\hat\phi)\psi -
ig\hbar\gamma_{i}\frac{\delta^2W}
 {\delta\bar\eta\delta J_{i}}.}
 \ea \ee Here we used the notations \be
 \hat\phi=\phi^{i}\gamma^{i},\qquad
 \phi^{i}=\{\sigma,\pi\},\qquad \gamma^{i}=\{1,i\gamma_5\}. \ee
 This system of equations may be solved by iterations with
 respect to either g or $\hbar$. But instead here we adopt the
 method used earlier for QED.

 The generating functional for this model can be written as \be
 \ba \Ds{ Z[J,\eta,\bar\eta]=\int D\phi D\bar\psi D\psi \exp
 i\left(L+J_{i}\phi^{i}+ \bar\psi\eta+\bar\eta\psi\right)=}\\ \\
\Ds{ =\int D\phi\exp\left(iS_{eff}+iJ_{i}\phi^{i}\right),} \ea
 \ee where we have integrated out the spinor variables and have
introduced the following notations \be \ba
 S_{eff}=-\frac{1}{2}\phi^2-F+i\hbar
 T,\qquad\phi^2=\sigma^2+\pi^2, \\ \\ F=\bar\eta\hat K\eta,
 \qquad T=\Tr\log\hat K, \qquad \hat K^{-1}=
 i\hat\partial+g\hat\phi. \ea \ee The free propagator for the
 auxiliary scalar field $\phi^{i} $ is \be D_{ij}(x-y)=\delta
 (x-y)\delta_{ij} \ee and $S_{eff}$ may be represented in the
 following form \be S_{eff}=-\frac{1}{2}\phi^i
 D_{ij}^{-1}\phi^j-F+i\hbar T.\ee Then for $\tilde\Gamma$ we have
 the following equation \be
  \frac{\delta\tilde\Gamma}{\delta\phi^i}=
 :\exp\left(\frac{i}{\hbar}\sum_{n=2}^{\infty}\frac{(-i\hbar)^n}{n!}
 G^{i_{1}\cdots
 i_{n}}\frac{\delta^n}{\delta\phi^{i_{1}}\cdots\delta
 \phi^{i_{n}}}\right)
 :\frac{\delta S_{eff}}{\delta\phi^{i}}. \ee Using the notations
 \be F_{i_{1}\cdots
 i_{n}}=(-1)^n\frac{\delta^n}{\delta\phi^{i_{1}}\cdots
 \delta\phi^{i_{n}}}F, \qquad
 T_{i_{1}\cdots
 i_{n}}=(-1)^n\frac{\delta^n}{\delta\phi^{i_{1}}
 \cdots\delta\phi^{i_{n}}}T,
 \ee we have for the derivative of $\tilde\Gamma$ \be \ba
 \Ds{\frac{\delta\tilde\Gamma}{\delta\phi^{i}}= -
 D^{-1}_{ij}\phi^j+ F_i-i\hbar T_i
-\frac{i}{2}\hbar G^{lj}\left(F_{ijl}
-i\hbar T_{ijl}\right)-} \\ \\ \Ds{
-\hbar^2\left[\frac{1}{6}G^{jkl}\left(-F_{lkji} +i\hbar
 T_{lkji}\right)-\frac{1}{8} G^{lj}G^{ks} \left( F_{jlksi}-i\hbar
 T_{jlksi}\right)\right]+\ldots.} \ea \ee Expanding the connected
 Green's functions $G^{ij}$ in terms of $\hbar$, we have \be \ba
 \Ds{\frac{\delta\tilde\Gamma}{\delta\phi^{i}}=
-D^{-1}_{ij}\phi^j+ F_i+\hbar
\left[-iT_i-\frac{i}{2}G^{lj}_{0}F_{jli}\right]+}\\ \\
\Ds{+\hbar^2\left[-\frac{i}{2}G_{1}^{lj}F_{jli}-
\frac{1}{2}G_{0}^{lj}T_{jli}+
\frac{1}{6}G_{0}^{jkl}F_{lkji}-\frac{1}{8}G_{0}^{jl}G_{0}^{ks}
F_{jlksi}\right]+\ldots.}
\ea \ee The reader should note that the equations for effective
action for the NJL model are analogous to those of QED, through
the substitutions $$\mu,\nu,\ldots,A_{\mu}\Rightarrow
i,j,\ldots,\phi^i.$$ For this reason we give expressions for the
effective action without carrying out the detailed calculations.

For $\tilde\Gamma_{0}$ we have \be
\frac{\delta\tilde\Gamma_{0}}{\delta\phi^i}=
-D^{-1}_{ij}\phi^j+ F_i \ee which gives us \be
\tilde\Gamma_{0}=-\frac{1}{2}\phi^i D_{ij}^{-1}\phi^j-F = S .\ee
Using Eq.(74) and $$
G_{0}^{lj}\frac{\delta^2\tilde\Gamma_{0}}{\delta\phi^j\delta\phi^k}=-
\delta^l_k $$
we set (in matrix notations) \be \hat G_{0}=\hat D\left(1+\hat
F\hat D\right)^{-1}, \ee where $\hat F\Rightarrow \{F_{ij}\} $
and $\hat D\Rightarrow \{D_{ij}\} $. Integration of the equation
\be \frac{\delta\tilde\Gamma_{1}}{\delta\phi^i}=
-iT_i-\frac{i}{2}G^{lj}_{0}F_{jli} \ee gives : \be
\tilde\Gamma_{1}=i\Tr\log \hat K+ \frac{i}{2}\Tr\log\left(1+\hat
F\hat D\right). \ee This is the total one--loop "effective
action" with fermionic part for the NJL model.

The equation for $\tilde\Gamma_{2}$ is: \be
\Ds{\frac{\delta\tilde\Gamma_{2}}{\delta\phi^i}=
-\frac{i}{2}G_{1}^{lj}F_{jli}- \frac{1}{2}G_{0}^{lj}T_{jli}+
\frac{1}{6}G_{0}^{jkl}F_{lkji}-\frac{1}{8}G_{0}^{jl}G_{0}^{ks}
F_{jlksi}.}
\ee Following the steps given in Appendix A for the case of the
QED we get: \be \tilde\Gamma_{2}=\frac{1}{2}\Tr\left(\hat
G_{0}\hat T\right)+
\frac{1}{8}G_{0}^{jl}G_{0}^{ks}F_{jlks}-\frac{1}{12}G_{0}^{js}
G_{0}^{kp}G_{0}^{ld}
F_{spd}F_{jkl}. \ee

The graphic representations for the $\tilde\Gamma_{2}$ are the
same as in QED excluding two points: \begin{itemize} \item Each
waveline is correspond to the Green's function of the
$\phi-$field--$D_{ij}$
      and the extra minus sign for each $D_{ij}$--line should be
take into account; \item It is necessary to take into account
that $D_{ij} $ is a $\delta-$function,
       i.e., two points in diagrams connected with $\phi-$lines
        must be reduced to one point in fact. \end{itemize}
\section{ The Two-loop Effective Potential for the NJL model in
the even--dimensional space--time}
We already pointed out that our expressions for $\tilde\Gamma$
are complicated integrals containing in integrands the classical
fields $A_{\mu},\bar\psi,\psi,\sigma,\pi$. It is possible to
integrate these expressions when these classical fields are set
equal to constants. In this case the effective action will be
transformed to the effective potential $V$: $$\Gamma=-\int dx
V.$$ The effective potential was calculated for many models
\ci{8} but usually only in the one-loop approximation. In this
section we would like calculate the effective potential for NJL
model in the two-loop approximation.

Omitting in the formulas Eqs.(75,77,79) the parts contributing to
fermion-fermion scattering the relevant part of the effective
action ( denoted below as $\tilde\Gamma_{\phi}$) for our purposes
is: \be \tilde\Gamma_{\phi}=\frac{1}{2}\phi^2+i\Tr\log\hat K+
\frac{1}{2}\Tr\left(\hat G_{0}\hat T\right) \ee (for the sake of
comparison with other authors we put here $\hbar=1 $). The terms
in Eq.(80) corresponds to the tree, the one-loop and the two-loop
actions, respectively. To return to the ordinary notations, it is
necessary to perform some calculations. Let's, for instance,
rewrite the second term as follows: \be \ba \Tr\log\hat
K=\Tr\log(i\hat\partial+g\hat\phi)^{-1}=\Tr\log((i\hat\partial)^{-1}
(1+g\hat\phi(i\hat\partial)^{-1})^{-1})=\\ \\
=\Tr\log\hat K_{0}-\Tr\log(1+g\hat\phi\hat K_{0}) \ea \ee and as
is usually done neglect the first term (because it gives us a
trivial infinite contribution). As shown in Appendix B for the
one -loop effective potential we have \be
\Ds{V_{1}=i\Tr\log(1+g\hat\phi\hat K_{0})=i2^{\frac{d}{2}-1}
\int\frac{d^dp}{(2\pi)^d}\log\left(1-\frac{g^2\phi^2}{p^2}\right),}
\ee
where $d$ is the dimension of the space-time.

For $d=2$ we get the following function \be \ba
\Ds{V_{1}=-\frac{1}{4\pi}\Lambda^2\log(1+g^2\phi^2/\Lambda^2)-
\frac{g^2\phi^2}{4\pi}
\log(1+\Lambda^2/g^2\phi^2)\simeq} \\ \\ \Ds{\simeq
-\frac{g^2\phi^2}{4\pi}\left[
1-\log(g^2\phi^2/\Lambda^2)\right],} \ea \ee where $\Lambda^2$ is
the cut-off parameter in the momentum space. The NJL model in two
dimensional space--time is renormalizable. We may renormalize our
effective potential $V$ in the two-dimensional space--time by
demanding \ci{1,8}: $$ \frac{\partial^2V}{\partial
\phi^2}\vert_{\phi_{0}} =1 ,$$ where $\phi^2_0 $ is arbitrary
subtraction point. Hence, the renormalised one--loop effective
potential is: $$
\Ds{V_r=\frac{1}{2}\phi^2+\frac{g^2\phi^2}{4\pi}\left(\log(\phi^2/
\phi_{0}^2)-3\right)}.$$

For $d=4$ we get from Eq.(80) the following one-loop formula: \be
\ba \Ds{V_{1}= -\frac{\Lambda^4}{16\pi^2}\ln
\left(1+\frac{g^2\phi^2}{\Lambda^2}\right)
-\frac{1}{16\pi^2}g^2\phi^2\Lambda^2+
\frac{g^4\phi^4}{16\pi^2}\log\left(1+\Lambda^2/g^2\phi^2\right)
\simeq }\\ \\ \Ds{\simeq -\frac{1}{8\pi^2}g^2\phi^2\Lambda^2+
\frac{g^4\phi^4}{16\pi^2}\left(\frac{1}{2}+
\log(\Lambda^2/g^2\phi^2)\right).} \ea \ee

Let's now turn to the calculation of the two-loop contribution to
the effective potential.

In the following expression \be \ba \frac{1}{2}\Tr\left(\hat
G_{0}\hat T\right)=\frac{1}{2}\left[\hat D(1+\hat F \hat
D)\right]^{ij}\Tr(\gamma_{j}\hat K\gamma_{i}\hat K)=\\ \\
=\frac{1}{2}g^2\Tr(\gamma_{j}\hat K\gamma_{i}\hat K)\delta^{ij} +
{\rm fermionic \quad parts}, \ea \ee we consider only the first
term and write it as: \be \int
dxV_{2}=-\frac{1}{2}g^2\Tr(\gamma_{j}\hat K\gamma_{i}\hat
K)\delta^{ij}=
-\frac{1}{2}g^2\int dx\Tr\left(\gamma_{i}\hat K(0)\gamma^{i}\hat
K(0)\right). \ee It is easy to see, that \be \hat
K(0)=\int[d^dp/(2\pi)^d](\hat
p+g\hat\phi)^{-1}=\sum_{n=0}^{\infty}(-g)^n \int
[d^dp/(2\pi)^d]\hat p (\hat\phi\hat p)^{n}(p^2)^{-n-1}. \ee
Taking into account the vanishing of the contributions of the
terms with even $ n $ and $$ \hat\phi\hat p\hat\phi\hat
p=\phi^2p^2 $$ we derive \be \hat K(0)=-g\int[d^dp/(2\pi)^d]\hat
p\hat\phi\hat p(p^2-g^2\phi^2)^{-1}(p^2)^{-1} \ee The
substitution of Eq.(88) into Eq.(86) gives us \be \ba
\Ds{V_{2}=-\frac{1}{2}g^4 \int[d^dp/(2\pi)^d]\int[d^dq/(2\pi)^d]
(p^2-g^2\phi^2)^{-1}(q^2-g^2\phi^2)^{-1}\cdot}\\ \\ \Ds{\cdot\Tr
\left(\gamma_{i}\hat p\hat\phi\hat p \gamma_{i}\hat q\hat\phi\hat
q \right).} \ea \ee Of course, one must understand integrals
 (87), (88), (89) as regularized, for example, by a cut-off
parameter $\Lambda^2$. Using the following identity: $$ \hat
p\hat\phi\hat p =(\sigma-i\gamma_{5}\pi)p^2 $$ we see that \be
\ba \gamma_{i}\hat p\hat\phi\hat p\gamma_{i}\hat q\hat\phi\hat q=
\gamma_{i}(\sigma-i\gamma_{5}\pi)\gamma_{i}(\sigma-i\gamma_{5}\pi)
p^2q^2=\\ \\
=p^2q^2[(\sigma-i\gamma_{5}\pi)+\sigma(i\gamma_{5})^2-i\gamma_{5}
(i\gamma_{5})^2
\pi](\sigma-i\gamma_{5}\pi)=0. \ea \ee Thus we see that the
two-loop contribution to the effective potential in the NJL model
vanishes and this result does not depend on the dimension of
space-time.

Concerning NJL model in four space-time dimension we see that as
indicated in the earlier work of Nambu and Jona-Lasinio there is
a relationship between $g^2$
     and the cut-off parameter $\Lambda^2$ (in the one-loop
     approximation ) \be g^2\Lambda^2 >4\pi^2,\ee whose validity
     is crucial for the dynamical generation of fermion mass.
     This relation can be made more precise. With this ultimate
     aim let us rewrite the four dimensional one-loop effective
     potential as follows \be
     V=\left(\frac{\Lambda^2}{2g^2}-\frac{\Lambda^4}{8\pi^2}\right)x
     +\frac{\Lambda^4}{16\pi^2}x^2\left(\frac{1}{2}-\log x\right)
\ee where we denoted $x=g^2\phi^2/\Lambda^2$. The minimum of this
function is obtained at \be x_{M}\log
 x_{M}=-1+4\pi^2/(g^2\Lambda^2) \ee which by taking into account
 of $x < 1$ (remembering, that $\Lambda^2$ is very large) gives
us Eq.(91). But for the second derivative of $V$ we have: $$ V"=-
\frac{\Lambda^4}{8\pi^2}(\log x+1), $$ from which we see that for
positivity at the point $ x_{M} $ we must have \be
4\pi^2/(g^2\Lambda^2) < 1-x_{M} .\ee
\section*{Acknowledgements}
One of the authors ( B.F.) would like to thank Prof. Namik K.Pak
for warm hospitality at T\"UBITAK and at the Midlle East
Technical University, Turkey. We also are gratefule to Prof.
Namik K.Pak for discussions , careful reading and corrections of
the manuscript .

This work was supported in part by the Foundation for Fundamental
Investigations of Uzbekistan State Committee for Science and
Technics under contract N 40, International Science Foundation
Grant MZJ000 and by DOPROG Program, T\"UBITAK, Turkey.
\appendix \section{Solving of the Eq.(60)}
 For making a transition from Eq.(60) to Eq.(61) we need to
calculate $G_1$ from the second of Eqs.(56): $$\Ds{
G_{1}^{\mu\nu}=G_0^{\mu\sigma}\frac{\delta^2\tilde\Gamma_1}
{\delta A^\sigma
\delta A^\lambda}G_0^{\lambda\nu}}$$ which gives us
$$\Ds{G_1^{\mu\nu}=G_0^{\mu\sigma}\left(iT_{\sigma\lambda}-
\frac{i}{2}\frac{\delta}
{\delta
A^\sigma}\left(G_0^{\alpha\beta}F^{\lambda}_{\beta\alpha}\right)
\right)G_0^{\lambda\nu}}.$$
Substituting the last equation into Eq.(60) we have $$ \ba
\Ds{\frac{\delta \tilde\Gamma_2}{\delta A_\mu}=-\frac{i}{2}
G_0^{\mu\sigma}\left(iT_{\sigma\lambda}-\frac{i}{2}\frac{\delta}
{\delta
A^\sigma}\left(G_0^{\alpha\beta}F^{\lambda}_{\beta\alpha}\right)\right)
  G_0^{\lambda\nu}F^\mu_{\sigma\lambda}- }\\ \\ \Ds{+\frac{1}{2}
G_0^{\sigma\lambda}T_{\sigma\lambda}^\mu
  +\frac{1}{6}G_0^{\sigma\lambda\nu}F_{\sigma\lambda\nu}^\mu
-\frac{1}{8}G_0^{\sigma\lambda}G_0^{\nu\rho}F_{\sigma\lambda\nu\rho}^\mu}
 \ea $$ The first and third terms of this relation can be
 combined as follows
 $$\Ds{-\frac{i}{2}G_0^{\mu\sigma}iT_{\sigma\lambda}G_0^{\lambda\nu}
 F^\mu_{\sigma\lambda}
 -\frac{1}{2} G_0^{\sigma\lambda}T_{\sigma\lambda}^\mu=
 \frac{1}{2}\frac{\delta}{\delta A_\mu}\Tr\left(\hat G_0 \hat
T\right)}\eqno(A.1)$$ where we have used the matrix notations: $
\hat G=\{G_{\mu\nu}\}, \hat T=\{T_{\mu\nu}\} $. Using for
simplicity the notation $\frac{\delta}{\delta
A_\mu}\Rightarrow\delta_\mu$ we have for remaining part of
$\delta\tilde\Gamma_2/\delta A_\mu$: $$ \ba
-\frac{1}{4}G_0^{\sigma\sigma_1}\delta_{\sigma_1}\left(G_0^{\alpha\beta}
\delta^{\lambda_1}F_{\beta\alpha}\right)G_0^{\lambda_1
\lambda}\delta^\mu F_{\sigma\lambda}
-\frac{1}{6}G_0^{\alpha\lambda}G_0^{\beta\nu}G_0^{\gamma\sigma}\delta^\mu
F_{\alpha\beta\gamma}F_{\lambda\nu\sigma}+\\ \\
+\frac{1}{8}G_0^{\alpha\beta} G_0^{\gamma\delta}\delta^\mu
F_{\alpha\beta\gamma\delta}. \ea $$ After some manipulations we
may transform the last expression into the following form $$
\Ds{\frac{\delta}{\delta A_\mu} \left[
  \frac{1}{8} G_0^{\alpha\beta} G_0^{\gamma\delta}
  F_{\alpha\beta\gamma\delta}
  -\frac{1}{12}G_0^{\alpha\sigma} G_0^{\beta\nu}
   G_0^{\gamma\lambda} F_{\alpha\beta\gamma}F_{\sigma\nu\lambda}
\right]}\eqno(A.2)$$ The sum of Eqs.(A.1) and (A.2) leads to the
Eq.(61).
\section{One--loop effective potential for the NJL model}
In this appendix we would like to compute the one-loop effective
potential for NJL model. All the steps of the calculation is
familiar and we present it only for completeness. The following
chain of equalities gives us the Eq.(82): $$ \ba \Ds{\int dx
V_{1}=i\Tr\log(1+g\hat\phi\hat
K_{0})=i\sum_{n=1}^{\infty}\frac{g^n}{n}(-1)^{n-1} \int
dx_{1}\cdots dx_{n}\cdot } \\ \\ \Ds{\cdot\Tr \left(\hat\phi\hat
K_{0}(x_{1}-x_{2}) \hat\phi\hat
K_{0}(x_{2}-x_{3})\cdots\hat\phi\hat K_{0}(x_{n}-x_{1})\right) =
}\\ \\ \Ds{=i\sum_{n=1}^{\infty}\frac{g^n}{n}(-1)^{n-1}\int dx dp
\Tr\left(\hat\phi\hat p^{-1}\hat\phi\hat p^{-1}\cdots\hat\phi\hat
p^{-1}\right)=}\\ \\
\Ds{=i\sum_{n=1}^{\infty}\frac{g^n}{n}(-1)^{n-1}\int dx dp
p^{-2n}\phi_{i_{1}} \phi_{i_{2}}\cdots\phi_{i_{n}}
\Tr\left(\gamma_{i_{1}}\hat p\gamma_{i_{2}}\hat
p\cdots\gamma_{i_{n}}\hat p\right)=}\\ \\
\Ds{=i2^{\frac{d}{2}}\int dx dp
\sum_{n=1}^{\infty}\frac{g^{2n}}{2n}(-1)^{2n-1}(\phi^2/p^2)^n=}\\
\\ \Ds{=i2^{\frac{d}{2}-1}\int dx dp \log(1-g^2\phi^2/p^2).} \ea
$$

\eject $ $ \vskip 10mm
\section*{Figures Captions} \vskip 15mm {\bf Figure 1.}
The two--loop contribution
to the effective action in the $\lambda\varphi^4$ theory. \vskip
15mm {\bf Figure 2.} Diagrams which corresponds to the term
$\Tr(\hat G_0 \hat T) $ in the $\tilde\Gamma_2 $. \vskip 15mm
{\bf Figure 3.} Diagrams which corresponds to the term $
G_0^{\alpha\beta} G_0^{\gamma\delta} F_{\alpha\beta\gamma\delta}
$ in the $\tilde\Gamma_2 $. \vskip 15mm {\bf Figure 4.} Diagrams
which corresponds to the term $ G_0^{\alpha\sigma} G_0^{\beta\nu}
G_0^{\gamma\lambda}
   F_{\alpha\beta\gamma}F_{\sigma\nu\lambda}$ \\in the
$\tilde\Gamma_2 $. \eject $ $ \vskip 70mm \centerline{a)} \vskip
90mm \centerline{b)} \vskip 10mm \centerline{Fig.1} \eject $ $
\vskip 40mm \centerline{a)} \vskip 130mm \centerline{b)} \vskip
10mm \centerline{Fig.2} \eject $ $ \vskip 90mm \centerline{Fig.3}
\vskip 90mm \centerline{Fig.4}
\end{document}